\documentclass[twocolumn]{IEEEtran}
\usepackage{amsmath,amssymb,graphicx,epsfig,cite,fancyhdr}

\newcommand{\be}{\begin{equation}}
\newcommand{\ee}{\end{equation}}
\newcommand{\ben}{\begin{equation*}}
\newcommand{\een}{\end{equation*}}
\newcommand{\ba}{\begin{eqnarray}}
\newcommand{\ea}{\end{eqnarray}}
\newcommand{\ban}{\begin{eqnarray*}}
\newcommand{\ean}{\end{eqnarray*}}
\newcommand{\bay}{\begin{array}}
\newcommand{\eay}{\end{array}}
\newcommand{\bat}{\begin{tabular}}
\newcommand{\eat}{\end{tabular}}
\newcommand{\bi}{\begin{itemize}}
\newcommand{\ei}{\end{itemize}}
\newcommand{\bin}{\begin{enumerate}}
\newcommand{\ein}{\end{enumerate}}
\newcommand{\bfig}{\begin{figure}}
\newcommand{\efig}{\end{figure}}
\newcommand{\tr}{\text{tr}}

\def\vs2{\vspace*{2mm}}

\title{
  MISO Capacity with Per-Antenna Power Constraint}
\author{Mai Vu \\
  Department of Electrical and Computer Engineering,
  McGill University, Montreal, H3A2A7\\
  Email: mai.h.vu@mcgill.ca}

\begin{document}
\pagestyle{fancy}
\renewcommand{\headrulewidth}{0pt} 
\lhead{\scriptsize } 
\rhead{\scriptsize \thepage}
\cfoot{}

\maketitle
\begin{abstract}
We establish in closed-form the capacity and the optimal signaling
scheme for a MISO channel with per-antenna power constraint. 
Two cases of channel state information are considered: constant
channel known at both the transmitter and receiver, and Rayleigh
fading channel known only at the receiver. For the first case, the
optimal signaling scheme is beamforming with the phases of the beam
weights matched to the phases of the channel coefficients, but the
amplitudes independent of the channel coefficients and dependent only
on the constrained powers. For the second case, the optimal scheme is
to send independent signals from the antennas with the constrained
powers. In both cases, the capacity with per-antenna power constraint
is usually less than that with sum power constraint.
\end{abstract}

\section{Introduction}
The capacity of a MIMO wireless channel depends on the constraints on
the transmit power and on the availability of the channel state
information at the transmitter and the receiver. With sum power
constraint across all transmit antennas, the capacity and the optimal
signaling are well established. For channels known at both the
transmitter and the receiver, the capacity can be obtained by
performing singular value decomposition of the channel and
water-filling power allocation on the channel eigenvalues
\cite{Cover_06}. For Rayleigh fading channels with coefficients known
only at the receiver, the ergodic capacity is obtained by sending
independent signals with equal power from all transmit antennas
\cite{Telatar_99}.

Under the per-antenna power constraint, the MIMO capacity is less
well understood. This per-antenna power constraint, however, is more
realistic in practice than sum power because of the constraint on the
individual RF chain connected to each antenna. Hence the transmitter
may not be able to allocate power arbitrarily among the transmit
antennas. Another appealing scenario for the per-antenna constraint is 
a distributed MIMO system, which has the transmitted antennas located
at different physical nodes that cannot share power with each
other. Thus understanding the  capacity and the optimal signaling
schemes under the per-antenna power constraint can be useful. 

The per-antenna power constraint has been investigated in different
problem setups. In \cite{yu2007transmitter}, the problem of a
multiuser downlink channel is considered with per-antenna power
constraint. It was argued that linear processing at both the
transmitter (by multi-mode beamforming) and the receiver (by MMSE 
receive beamforming with successive interference cancellation) can
achieve the capacity region. Using uplink-downlink duality, the
boundary points of the capacity region for the downlink channel with
per-antenna constraint can be found by solving a dual uplink problem,
which maximizes a weighted sum rate for the uplink channel with sum
power constraint across the users and an uncertain noise. The dual
uplink problem is convex which facilitates computation. In
\cite{shi2008perantenna}, an iterative algorithm based on geometric
programming is proposed for maximizing the weighted sum rate of
multiple users with per-antenna power constraint. In 
\cite{codreanu2007mimo}, another iterative method is proposed for
solving the sum rate maximization problem under the more generalized
power constraints on different groups of antennas. However, in all of
these works, because of the complexity of the optimization problem, no
closed-form analytical solutions of the optimal linear transmit
processing scheme or the capacity were proposed. To the best of our
knowledge, such closed-form solutions (for a MIMO channel with
per-antenna power constraint) are not available even in the
single-user case.

In this letter, we establish in closed-form the capacity and optimal
signaling scheme for the single-user MISO channel with per-antenna
power constraint. In this channel, the transmitter has multiple
antennas and the receiver has a single antenna. Both cases of constant
channel known to both the transmitter and receiver and of Rayleigh
fading channel known only to the receiver are considered. When the
channel is constant and known at both the transmitter and receiver, it
turns out that the capacity optimal scheme is single-mode beamforming
with the beam weights matched to the channel phases but not the
channel amplitudes. Our result covers the special case of 2 transmit
antennas considered in \cite{liu2006capacity} as part of a Gaussian
multiple access channel channels with common data, in which it was
established that beamforming of the common data only maximizes the sum
rate, which is equivalent to beamforming in a MISO channel. When the
channel is Rayleigh fading and is known only to the receiver, the
optimal scheme is sending independent signals from the transmit
antennas with the constrained powers. In both cases, the capacity with
per-antenna power constraint is usually less than that with sum power
constraint.

In establishing these results, we need to solve the corresponding
capacity optimization problems. In the constant channel case, our
proof method is to solve a relaxed problem and then show that the
solution of the relaxed problem satisfies the original constraints and
hence is optimal. In the fading channel case, our proof is based on
the symmetry of the Raleigh fading distribution. This latter technique
can be generalized directly to the MIMO fading channel with
per-antenna power constraint.

The rest of this letter is organized as follows. In Section
\ref{model}, we discuss the MISO channel model, the capacity
optimization problem and the different power constraints. Then the
results for constant channels known at both the transmitter and
receiver are established in Section \ref{constant_channel}, and for
Rayleigh fading channels known only at the receiver in Section
\ref{fading_channel}. In Section \ref{conclusion}, we provide some
concluding remarks. For notation, we use bold face lower-case letters
for vectors, capital letters for matrices, $(.)^T$ for transpose, 
$(.)^\ast$ for conjugate, $(.)^\dagger$ for conjugate transpose,
$\succcurlyeq$ for matrix inequality (positive semi-definite
relation), $\tr(.)$ for trace, and ${\text{diag}}\{.\}$ for forming a 
diagonal matrix with the specified elements.

\section{Channel Model and Power Constraints}
\label{model}

\subsection{Channel model}
Consider a multiple-input single-output (MISO) channel with $n$
transmit antennas. Assuming flat-fading, the channel from each antenna
is a complex, multiplicative factor $h_i$. Denote the channel
coefficient vector as ${\mathbf h} = [h_1 \ldots h_n]^T$, and the
transmit signal vector as ${\bf x} = [x_1 \ldots x_n]^T$. Then the
received signal can be written as
\be
  y = {\bf h}^T {\bf x} + z
  \label{miso_channel}
\ee
where $z$ is a scalar additive white complex Gaussian noise with power
$\sigma^2$.

We assume that the channel coefficient vector ${\bf h}$ is
known at the receiver, which is commonly the case in practice with
sufficient receiver channel estimation. We consider 2 cases of
channel information at the transmitter: constant channel coefficients
also known to the transmitter, and fading channel coefficients which
are circularly complex Gaussian and are not known to the
transmitter. The former can correspond to a slow fading environment,
whereas the latter applies to fast fading.

The capacity of this channel depends on the power constraint on the
input signal vector ${\bf x}$. In all cases, however, because of the
Gaussian noise and known channel at the receiver, the optimal input
signal is Gaussian with zero mean \cite{Telatar_99}. Let ${\bf Q} =
E[{\bf x x}^\dagger]$ be the covariance of the Gaussian input, then
the achievable transmission rate is 
\be
  R = \log\left(1 + \frac{1}{\sigma^2} {\bf h}^T {\bf Q} {\bf h}^\ast \right).
\ee
The remaining question is to establish the optimal ${\bf Q}$ that maximizes
this rate according to a given power constraint.

\subsection{Power constraints}
Often the MISO channel is studied with sum power constraint across 
all antennas. In this letter, we study a more realistic per-antenna
power constraint. For comparison, we also include the case of
independent multiple-access power constraint. We elaborate on each
power constraint below.

\subsubsection{Sum power constraint}
With sum power constraint, the total transmit power from all $n$
antennas is $P$, but this power can be shared or allocated arbitrarily
among the transmit antennas. This constraint translates to the
condition on the input covariance as $\tr({\bf Q}) \leq P$.

\subsubsection{Independent multiple-access power constraint}
In this case, each transmit antenna has its own power budget and acts
independently. This constraint can model the case of distributed
transmit antennas, such as on different sensing nodes scattered in a
field, without explicit cooperation (in terms of coding and signal 
design) among them. Let $P_i$ be the power constraint on antenna $i$,
then this constraint is equivalent to having a diagonal input
covariance ${\bf Q} = {\text{diag}}\{P_1, \ldots, P_n\}$.

\subsubsection{Per-antenna power constraint}
Here each antenna has a separate transmit power budget of $P_i$ ($i =
1,\ldots,n$) and can fully cooperate with each other. Such a channel
can model a physically centralized MISO system, for example, the
downlink of a system with multiple antennas at the basestation and
single antenna at each user. In such a centralized system, the
per-antenna power constraint comes from the realistic individual
constraint of each transmit RF chain. The channel can also model a
distributed (but cooperative) MISO system, in which each transmit
antenna belongs to a sensor or ad hoc node distributed in a
network. In such a distributed scenario, the nodes have no ability to
share or allocate power among themselves and hence the per-antenna 
power constraint holds (but they may wish to cooperate to design codes
and transmit signals). The per-antenna constraint is equivalent to
having the input covariance matrix ${\bf Q}$ with fixed diagonal values
$q_{ii} = P_i$. Note that this constraint is on the diagonal values of 
${\bf Q}$ and is not the same as having the eigenvalues of ${\bf Q}$
equal to $P_i$.

\section{MISO capacities with constant channels}
\label{constant_channel}
In this section, we investigate the case that the channel is constant and
known at both the transmitter and the receiver. First, we briefly
review known results on the capacity of the channel in
(\ref{miso_channel}) under sum power constraint and independent
multiple access constraint. Then we develop the new result on MISO
capacity with per-antenna power constraint.

\subsection{Review of known capacity results}
\label{known_cap_results}
\subsubsection{MISO capacity under sum power constraint}
With sum power constraint, the capacity optimization problem can be posed as
\ba
  \max && \log\left(1 + \frac{1}{\sigma^2} {\bf h}^T {\bf Q} {\bf h}^\ast \right)
  \label{cap_sumpwr}\\
  \text{s.t.} && \tr({\bf Q}) \leq P \;, \quad {\bf Q} \succcurlyeq 0 \nonumber
\ea
where ${\bf Q}$ is Hermitian. This problem is convex in ${\bf Q}$. Let
${\bf Q} = {\bf U\Lambda U}^\dagger$ be the eigenvalue decomposition,
then the optimal solution is to pick an eigenvector ${\bf u}_1 = {\bf
  h}^\ast/\|{\bf h}\|$ and allocate all transmit power in this
direction, that is, the first eigenvalue $\lambda_1 = P$.

Thus the transmitter performs single-mode beamforming with the optimal
beam weights as ${\bf h}^\ast/\|{\bf h}\|$. At each time, all transmit
antennas send the same symbol weighted by a specific complex weight at
each antenna. In this optimal beamforming, the beam weight on an
antenna not only has the phase matched to (being the negative of) the
phase of the channel coefficient from that antenna, but also the
amplitude proportional to the amplitude of that channel
coefficient. In other words, power is allocated among the antennas
proportionally to the channel gains from these antennas.

The MISO capacity with sum power constraint is
\be
  C_s = \log\left(1 + \frac{P}{\sigma^2} \sum_{i=1}^n |h_i|^2
  \right) = \log\left(1 + \frac{P}{\sigma^2} \|{\bf h}\|^2 \right).
  \label{sum_cap_sumpwr}
\ee

\subsubsection{Independent multiple-access capacity}
Under the independent multiple-access constraint, the capacity is 
equivalent to the sum capacity a multiple access channel, without
explicit cooperation among the transmitters, as \cite{Cover_06}
\begin{align}
  C_{ma} = \log\left( 1+ \frac{1}{\sigma^2} \sum_i P_i|h_i|^2 \right).
  \label{mac_sum_rate}
\end{align}
In this case, there is no optimization since ${\bf Q} =
\text{diag}\{P_1, \ldots, P_n\}$. The transmit antennas send different
and independent symbols at each time.

\subsection{MISO capacity with per-antenna power constraint}
\label{MISO_peak_cap}
The capacity with per-antenna power constraint can be found by solving
the following optimization problem:
\ba
  \max && \log\left(1 + \frac{1}{\sigma^2} {\bf h}^T {\bf Q} {\bf h}^\ast \right)
  \label{cap_n_peakpwr}\\
  \text{s.t.} && q_{ii} \leq P_i \quad i = 1, 2, \ldots, n \nonumber \\
  && {\bf Q} \succcurlyeq 0. \nonumber
\ea
Noting that the per-antenna power constraint $q_{ii} \leq P_i$ can be
written as ${\bf e}_i^T {\bf Q} {\bf e}_i \leq P_i$ where ${\bf e}_i
= [0 \ldots 1 \ldots 0]^T$ is a vector with the $i^\text{th}$ element
equal to $1$ and the rest is $0$, thus this constraint is linear in
${\bf Q}$. Thus the above problem is also convex. So far, however,
there is no closed-form solution available.

We are able to solve the above problem (\ref{cap_n_peakpwr})
analytically with closed-form solution by first applying a matrix
minor condition to relax the positive semi-definite constraint ${\bf Q}
\succcurlyeq 0$, reducing the problem to a form solvable in
closed-form, and then showing that the optimal solution to the relaxed
problem is also the optimal solution to the original problem. The
details are given in Appendix \ref{app_const}.

It is also possible to show that the optimal covariance of
\eqref{cap_n_peakpwr} has the rank satisfying $\text{rank}({\bf
  Q}^\star) \leq \text{rank}({\bf h})$. Hence for the MISO channel
considered here, $\text{rank}({\bf Q}^\star) = 1$ and the optimal
signaling is beamforming. This proof is provided in Appendix
\ref{app_rank}.

Here we describe the optimal covariance ${\bf Q}^\star$ and discuss the
meaning of the solution. The optimal ${\bf Q}^\star$ has the elements given
as
\be
  q_{ij} = \frac{h_i^\ast h_j}{|h_ih_j|}{\sqrt{P_i P_j}} \;, \quad i,j =
  1, \ldots, n. 
  \label{opt_qij}
\ee
Let ${\bf Q}^\star = {\bf V \Lambda V}^\dagger$ be its eigenvalue 
decomposition. ${\bf Q}^\star$ can be shown to have rank-one with the
single non-zero eigenvalue as $\lambda_1 = \sum P_i$ and the
corresponding eigenvector ${\bf v}_1$ with elements given as 
\be
  v_{k1} = \frac{h_k^\ast}{|h_k|} \frac{\sqrt{P_k}}{\sqrt{P}} = \eta_k
  \frac{\sqrt{P_k}}{\sqrt{P}} 
  \label{opt_eigvec_element}
\ee
where $P = \sum P_i$ and $\eta_k = h_k^\ast/|h_k|$ is a point on the
complex unit circle with phase as the negative of the phase of $h_k$.

The optimal signaling solution with per-antenna power constraint
is beamforming with the beam weight vector as 
${\bf v}_1^\ast$. Different from the sum power constraint case, here,
the beam weight only has the phase matched to the phase of the channel 
coefficient, but the amplitude independent of the channel and fixed 
according to the power constraint. Thus there is {\em no power
  allocation} among the transmit antennas: the transmit power from the
$i^{th}$ antenna is fixed as $P_i$.

For beamforming, it is useful to examine the angle $\theta$ ($0 \leq
\theta \leq \pi/2$) between a beam weight vector ${\bf w}$ and the
channel vector ${\mathbf h}$, defined as $\cos \theta = {\bf
  h}^\dagger {\bf w}/(\|{\bf h}\| \cdot \|{\bf w}\|)$. This angle
$\theta$ affects the capacity as follows:
\begin{align}
   C_p & = \log\left(1 + \frac{1}{\sigma^2} P\|{\bf h}\|^2
     \cos\theta \right) .
\end{align}
Hence the smaller the angle, the larger the capacity. As in the case
with sum-power constraint, the beam weight ${\bf w} = {\bf
  h}^\ast/\|{\bf h}\|$ completely matches the channel (both the phase
and amplitude) and the capacity as obtained in (\ref{sum_cap_sumpwr})
is the maximum.

With per-antenna power constraint, the beam-weight vector is ${\bf w} =
{\bf v}_1$ and the angle $\theta$ satisfies
\be
   \cos \theta = \sum_k \frac{h_k}{|h_k|} 
   \frac{\sqrt{P_k}}{\sqrt{P}} \frac{h_k^\ast}{|h|} =
   \frac{1}{|h|\sqrt{P}} \sum_k |h_k| \sqrt{P_k}.
   \label{theta}
\ee
Applying the Cauchy-Schwartz's inequality on (\ref{theta}), the
maximum $\cos\theta=1$ occurs if and only if $\sqrt{P_k} = c |h_k|$
for some constant $c$ and for all $k=1,\ldots, n$. In all other cases,
$\cos\theta < 1$ and hence $\theta > 0$. Thus with the per-antenna
power constraint, except for the special case in which the power
constraints $P_i$ happens to be proportional to the channel
coefficient amplitude $|h_i|$, the beamforming vector ${\bf v}_1^\ast$
does not completely align with the channel vector ${\bf
  h}$. Nevertheless, it provides the largest transmission rate without 
power allocation.

Our result also covers the case of 2 user multiple access channel
with common data considered in \cite{liu2006capacity}, which states
that the sum rate is maximized by just sending the common data and 
performing beamforming.

The MISO capacity with per-antenna power constraint is
\be
  C_p = \log\left[1 + \frac{1}{\sigma^2}\left(\sum_{i=1}^n
      |h_i|\sqrt{P_i} \right)^2 \right].
  \label{MISO_cap_peakpwr_n}
\ee
Compared to (\ref{sum_cap_sumpwr}) and (\ref{mac_sum_rate}), we see
that $C_{ma} \leq C_p \leq C_s$.

\subsection{Numerical examples}
\subsubsection{With 2 transmit antennas}
We provide numerical examples of the capacities for a MISO channel
with 2 transmit antennas. Assume a complex test channel ${\bf h} =
[0.3 + 0.2i \;\; 0.4 - 0.7i]^T$. For fair comparison, the total
transmit power in the sum power constraint must equal the sum of the
individual powers in the per-antenna power constraint. Thus we choose
the transmit powers such that $P_1+P_2=P=10$.

\begin{figure}[!tb]
  \centering{\epsfig{figure=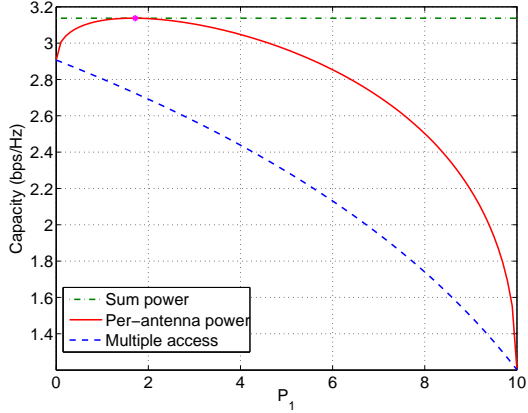, width=7cm}}
  \caption{Capacities for a $2\times 1$ constant channel under different
    power constraints.}
  \label{mac2_sumcap}
\end{figure}

Figure \ref{mac2_sumcap} shows the MISO capacity versus $P_1$ under
the three different power constraints: sum power constraint
(\ref{sum_cap_sumpwr}), independent multiple-access power constraint
(\ref{mac_sum_rate}), and per-antenna power constraint
(\ref{MISO_cap_peakpwr_n}). Compared to the multiple access capacity
which is obtained with independent signals from the different transmit
antennas, we see that introducing correlation among the transmit
signals by beamforming increases the capacity. (Single-mode beamforming
introduces complete correlation among the signals from different antennas
since all antennas send the same symbol, just with different weights.)
Under the sum power constraint, power allocation can further increase
the capacity.

The two MISO capacities with sum power constraint and per-antenna
power constraint are equal at a single point when the value of $P_1$
is such that $P_1/P_2 = |h_1|^2/|h_2|^2$, which is $P_1^\star=1.72$ in 
this example. On the other hand, at the equal power point $P_1=P_2=5$, 
the capacity with per-antenna power constraint is about 93\% of that
with sum power constraint, and is almost 30\% higher than the multiple
access capacity.

\subsubsection{With $n$ transmit antennas}
\begin{figure}[!tb]
  \centering{\epsfig{figure=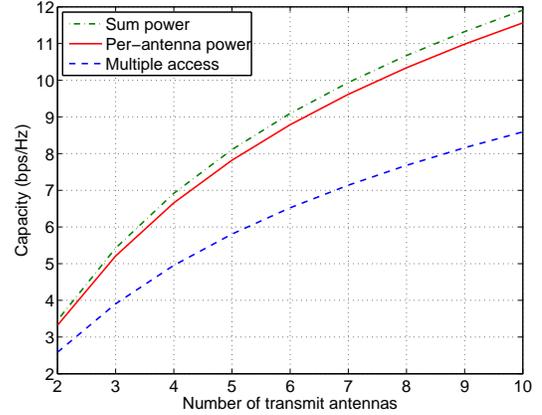, width=7cm}}
  \caption{Sum capacity for the power symmetric case with $h_k = k$.}
  \label{macn_pwrsym}
\end{figure}

With $n$ transmit antennas ($n > 2$), it is more informative to
study the capacity as a function of $n$. To make some insightful
comparisons, we consider the case in which all users have 
the same transmit power budget $P_i=P_0=1$ and $\sigma^2=1$. We can
see that if the channel is also symmetric ($h_i=h_j$ for all $i,j$)
then the capacity with per-antenna power constraint is the same as
that with sum power constraint. Now suppose that the channel is
non-symmetric as $h_k=k$. Then the 3 capacities become
\begin{align*}
  {\cal C}_{ma} &= \log\left( 1+ \frac{P_0}{\sigma^2}
    \left(\frac{n^3}{3}+\frac{n^2}{2} + \frac{n}{6} \right) \right) \\
  {\cal C}_p &= \log\left( 1+ \frac{P_0}{\sigma^2}
    \frac{(n^2+n)^2}{4} \right) \\
  {\cal C}_s &= \log\left( 1+ \frac{n P_0}{\sigma^2}
    \left(\frac{n^3}{3}+\frac{n^2}{2} + \frac{n}{6} \right) \right).
\end{align*}
Figure \ref{macn_pwrsym} shows these capacities versus the number
of users $n$. In this case, the capacity with per-antenna power
constraint is almost as high as the capacity with sum power
constraint, and both are significantly better than the multiple 
access capacity.

\section{MISO capacities with fading channels}
\label{fading_channel}

In this section, we study Rayleigh fading channels, in which the
channel coefficients $h_i$ are now independent, zero-mean complex
circularly Gaussian random variables with unit variance. We assume
that the channel vector ${\bf h}$ is known perfectly to the receiver
but is unknown to the transmitter.

Again we first review the known capacity results with sum power
constraint and independent multiple access power constraint, then
establish the new result with per-antenna power constraint.

\subsection{Review of known capacity results}
\subsubsection{MISO capacity with sum power constraint}
For a MISO fading channel with sum power constraint, the capacity
is a special case of \cite{Telatar_99}. The optimal covariance of the
Gaussian transmit signal is ${\bf Q}= \frac{P}{n}{\bf I}$, implying that
each antenna sends independent signal with equal power. The ergodic
MISO capacity is 
\be
  {\cal C}_s = E_{{\bf h}} \left[ \log\left(1 +
      \frac{P}{n\sigma^2} \|{\bf h}\|^2 \right) \right].
  \label{erg_sum_cap_sumpwr}
\ee
Compared to (\ref{sum_cap_sumpwr}), there is a dividing factor of $n$
in the power in the instantaneous capacity equation. This power loss 
factor is due to the lack of channel information at the transmitter.

\subsubsection{Independent multiple-access capacity}
In this case, the transmit covariance is ${\bf Q} = \text{diag}\{P_1,
\ldots, P_n\}$. The capacity is obtained by averaging the instantaneous
capacity in (\ref{mac_sum_rate}) over fading
\cite{shamai1997macfading}. Specifically, the ergodic capacity is  
\be
  {\cal C}_{ma} = E_{{\bf h}} \left[ \log\left( 1+
    \frac{1}{\sigma^2} \sum_{i=1}^n P_i|h_i|^2 \right) \right].
  \label{erg_sum_cap_mac}
\ee

\subsection{MISO capacity with per-antenna power constraint}
To establish the ergodic MISO capacity with per-antenna power
constraints, we need to solve the following stochastic version of
problem (\ref{cap_n_peakpwr}):
\ba
  \max && E_{{\bf h}} \left[ \log\left(1 + \frac{1}{\sigma^2} {\bf
        h}^T {\bf Q} {\bf h}^\ast \right) \right]
  \label{erg_cap_peakpwr}\\
  \text{s.t.} && {\bf Q} \succcurlyeq 0 \;, \quad q_{ii} \leq P_i \;,
  \quad i = 1, \ldots, n, \nonumber
\ea
where ${\bf Q}$ is Hermitian.

Since the per-antenna constraint $q_{ii} \leq P_i$ is not the same as
a constraint on the eigenvalues of ${\bf Q}$, the analysis for fading
channels as in \cite{Telatar_99} cannot be applied here. That is, if
we perform the eigenvalue decomposition ${\bf Q} = {\bf U}_Q 
{\bf \Lambda}_Q {\bf U}_Q^\dagger$, then although ${\bf h}^T{\bf U}_Q$
has the same distribution as ${\bf h}^T$, the diagonal values of
${\bf \Lambda}_Q$ do not have the same constraints as the diagonal
values of ${\bf Q}$. Hence the problem is no longer equivalent through
eigen-decomposition.

However, by also relying on the centrality and symmetry of the
Rayleigh fading distribution in a slightly different way, we show that
the optimal solution of (\ref{erg_cap_peakpwr}) is ${\bf Q} = 
\text{diag}\{P_1, \ldots, P_n\}$. The details are given in Appendix
\ref{app_fading}.

The optimal solution means that each transmit antenna sends
independent signal at its full power. Somewhat surprisingly, this is  
the same transmit strategy under the independent multiple access
constraint. Hence the ergodic capacity with per-antenna power
constraint is
\be
  {\cal C}_p = {\cal C}_{ma}.
  \label{erg_sum_cap_peak}
\ee
Thus, for a Rayleigh fading without channel information at the
transmitter, having the possibility for cooperation among the transmit 
antennas under the per-antenna power constraint does not increase the
average capacity. (It should be noted, however, that cooperation
without transmit channel state information can still increase
reliability significantly \cite{Laneman_03}.)

From (\ref{erg_sum_cap_sumpwr}), (\ref{erg_sum_cap_mac}) and
(\ref{erg_sum_cap_peak}), we can show that
\be
  {\cal C}_{p} \leq {\cal C}_s
\ee
always holds. This is proven by noticing that the channel
coefficients $h_i$ are i.i.d. Thus for any permutation $\pi =
\text{perm}(1, \ldots, n)$, we can express ${\cal C}_{p}$ as
\ben
  {\cal C}_{p} = E_{{\bf h}} \left[ \log\left( 1+
      \frac{1}{\sigma^2} \sum_{i=1}^n P_i|h_{\pi_i}|^2 \right) \right].
\een
Let $\pi^{(k)} = (k, \ldots, n, 1, \ldots, k-1)$ which is a rotation
of the order. Then based on the concavity of the $\log$ function, the
following expressions holds:
\begin{align*}
  {\cal C}_{p} & = \frac{1}{n} \sum_{k=1}^n E_{{\bf h}} \left[
    \log\left( 1+ \frac{1}{\sigma^2} \sum_{i=1}^n P_i|h_{\pi^{(k)}_i}|^2
    \right) \right] \\
  & \leq E_{{\bf h}} \left[ \log\left( 1+ \frac{1}{n} \sum_{k=1}^n
      \frac{1}{\sigma^2} \sum_{i=1}^n P_i|h_{\pi^{(k)}_i}|^2  
    \right) \right] = {\cal C}_{s}.
\end{align*}
Equality holds if and only if $P_i=P/n$ for all $i$.

We see that similar to the case of sum power constraint, under the
per-antenna power constraint, the presence or lack of channel
information at the transmitter has a significant impact on the optimal
transmit strategy and the channel capacity. With full channel state 
information, the optimal strategy under either power constraint is
beamforming (sending completely correlated signals), while without
channel state information at the transmitter, the optimal strategy is
to send independent signals from the different antennas.

\subsection{Numerical examples}
\begin{figure}[!tb]
  \centering{\epsfig{figure=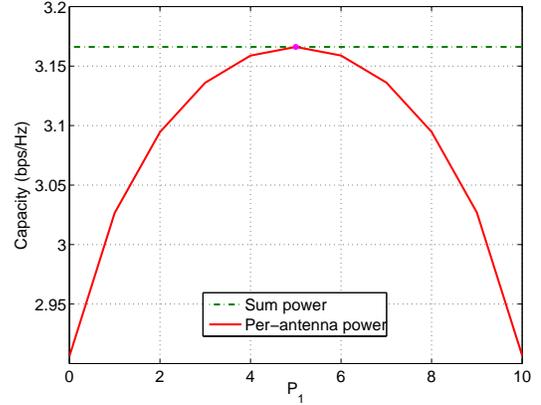, width=7cm}}
  \caption{Ergodic capacities for a $2 \times 1$ Rayleigh fading
    channel under different power constraints.}
  \label{mac2_fading_C0P1}
\end{figure}
For numerical example, we examine a MISO fading channel with 2
transmit antennas. Figure \ref{mac2_fading_C0P1} shows the plots of
the ergodic capacities in (\ref{erg_sum_cap_sumpwr}),
(\ref{erg_sum_cap_mac}) and (\ref{erg_sum_cap_peak}) versus the
transmit power constraint on the first antenna. The symmetry observed
in these plots is a result of the average over fading. The
difference in the ergodic capacities with sum power constraint and
with per-antenna power constraint are smaller in fading channels than
in constant channels. The two capacities are equal at the point
$P_1=P/2=5$.

\section{Conclusion}
\label{conclusion}

We have established the MISO capacity with per-antenna power
constraint for 2 cases of channel state information. In the case of
constant channel known to both the transmitter and the receiver, the
capacity is obtained by beamforming. The optimal beam weights,
however, are different from those under the sum power
constraint. Specifically, only the phases of the beam weights are
matched to the phases of the channel coefficients, but the amplitudes
are independent of the channel and depend only on the constrained
powers. In the case of Rayleigh fading channel known to the receiver
only, the capacity is obtained by sending independent signals from the
transmit antennas with the constrained powers. In both cases, the
capacity with per-antenna power constraint is usually less than that
with sum power constraint.

Our proof technique for the case of Rayleigh fading channel can be
applied directly to the more general MIMO fading channel with
per-antenna power constraint. For the case of constant channel known
at both the transmitter and receiver, however, the proof technique
here may not be generalized directly to the MIMO channel, except that
of the rank. The capacity of a constant MIMO channel with per-antenna 
constraint is still an open problem.

\section*{Acknowledgment}
The author would like to thank an anonymous reviewer for pointing out 
the result and proof (in Appendix \ref{app_rank}) for the rank of the
optimal covariance matrix for constant channels. Thanks also to Long
Gao from Hitachi Corporation for pointing out that the per-antenna
constraint is linear in ${\bf Q}$ and hence the capacity optimization
problem is also convex.

\appendix
\subsection{Optimal transmit covariance for constant channels}
\label{app_const}
In problem (\ref{cap_n_peakpwr}), the optimal ${\bf Q}^\star$ must have
the diagonal values $q_{ii} = P_i$, for otherwise, we can singularly
increase the diagonal value of ${\bf Q}$ that is less than its corresponding
power constraint and hence increase the objective function. The
problem remains to find the off-diagonal entries $q_{ij}$ ($i \neq j$).

The main difficulty here is the semi-definiteness constraint ${\bf Q}
\succcurlyeq 0$. This constraint is equivalent to having all principal
minors of ${\bf Q}$ being positive semi-definite \cite{Prussing_86}. Thus
the constraint involves multiple polynomial constraints on $q_{ij}$
with degree up to $n$.

To solve this problem, we consider a relaxed version with semi-definite
constraints involving only $2 \times 2$ principal minors of ${\bf Q}$ of the
form
\be
  M_{(ij)} = \left[
    \begin{array}{cc}
      P_i  &  q_{ij}^\ast \\
      q_{ij} & P_j
    \end{array}
 \right].
\ee
Such a minor is obtained by removing $n-2$ columns (except columns $i$
and $j$) and the correspondingly transposed $n-2$ rows of ${\bf Q}$. We then
form the following relaxed problem:
\ba
  \max && {\bf h}^T {\bf Q} {\bf h}^\ast 
  \label{cap_n_peakpwr_relaxed}\\
  \text{s.t.} && q_{ii} = P_i , \quad i = 1, 2, \ldots, n \nonumber \\
  && M_{(ij)} \succcurlyeq 0 , \quad i \neq j,\; i,j = 1, 2, \ldots, n \nonumber.
\ea
Since this problem is a relaxed version of (\ref{cap_n_peakpwr}), if
the optimal ${\bf Q}^\star$ of this relaxed problem is positive
semi-definite, then it is also the optimal solution of
(\ref{cap_n_peakpwr}).

The constraint $M_{(ij)} \succcurlyeq 0$ is equivalent to $|q_{ij}|^2
\leq P_i P_j$. Based on this, we can form the Lagrangian as 
\ben
  {\cal L}(q_{ij}, \lambda_{ij}) = {\bf h}^T {\bf Q} {\bf h}^\ast - \sum_{i \neq j}
  \lambda_{ij} \left( |q_{ij}|^2 - P_iP_j \right), 
\een
where $\lambda_{ij}$ are the Lagrange multipliers. Differentiate ${\cal
  L}$ with respect to $q_{ij}$ (for the differentiation of a real
function with respect to a complex variable, we use the rules of
Wirtinger calculus as discussed in \cite{Fischer_02}, Appendix A) to 
get
\ben
  \frac{\partial \cal L}{\partial q_{ij}} = h_i^\ast h_j -
  \lambda_{ij} q_{ij}.
\een
Equating this expression to zero, we have
\be
  q_{ij} = \frac{h_i^\ast h_j}{\lambda_{ij}}.
  \label{q_lambda}
\ee
The optimal $q_{ij}$ should satisfy its constraint with equality, that
is $|q_{ij}|^2 = P_i P_j$. This is because the terms that contain
$q_{ij}$ in the objective function are $q_{ij} h_i h_j^\ast +
q_{ij}^\ast h_i^\ast h_j$. Thus if $|q_{ij}|^2 < P_i P_j$, we can
increase $q_{ij}$ by a real amount $\Delta_{ij}$ with the same sign as
the sign of $h_i h_j^\ast + h_i^\ast h_j$, resulting in a positive
increase in the objective function.

Combining (\ref{q_lambda}) and $|q_{ij}|^2 = P_i P_j$, we have
$\lambda_{ij} = |h_ih_j|/\sqrt{P_i P_j}$, which leads to the optimal
value for $q_{ij}$ as given in (\ref{opt_qij}). Since $\lambda_{ij} >
0$, a simple check on the second derivative of ${\cal L}$ shows that
this $q_{ij}$ is the maximum point of the relaxed problem
(\ref{cap_n_peakpwr_relaxed}).

The resulting covariance matrix ${\bf Q}^\star$ is indeed positive
semi-definite. It has a single positive eigenvalue as $\lambda_1 =
\sum P_i$ and $n-1$ zero eigenvalues. Therefore it is also the optimal
solution of problem (\ref{cap_n_peakpwr}).

The eigenvector corresponding to the non-zero eigenvalue of ${\bf Q}^\star$
has the elements given by (\ref{opt_eigvec_element}).

\subsection{The rank of the optimal transmit covariance for constant channels}
\label{app_rank}
Consider problem \eqref{cap_n_peakpwr} and rewrite it as
\begin{eqnarray*}
  \max && \log\left(1 + \frac{1}{\sigma^2} {\bf h}^T {\bf Q} {\bf h}^\ast \right)\\
  \text{s.t.} && {\bf e}_i^T {\bf Q} {\bf e}_i \leq P_i , \quad i =
  1\ldots n \\
  && {\bf Q} \succcurlyeq 0. \nonumber
\end{eqnarray*}
where ${\bf e}_i = [0 \ldots 1 \ldots 0]^T$ is a vector with the
$i^\text{th}$ element equal to $1$ and the rest is $0$. Since this
problem is convex ${\bf Q}$, Lagrangian method can be used to obtain
the exact solution.

Denote ${\bf P}=\text{diag}\{P_i\}$, ${\bf D}=\text{diag}\{\lambda_{i}\}$
as a diagonal matrix consisting of Lagrangian multipliers for the 
per-antenna power constraints, and ${\bf M} \succcurlyeq 0$ as the
Lagrangian multiplier for the positive semi-definite constraint. We
can then form the Lagrangian as
\begin{equation*}
  {\cal L} = {\bf h}^T {\bf Q} {\bf h}^\ast - \tr[ {\bf D}({\bf Q} - 
    {\bf P})] + \tr({\bf M} {\bf Q}).
\end{equation*}
Taking its first order derivative with respect to ${\bf Q}$ (see
\cite{van2002optimum} Appendix A.7 for derivatives with respect to a
matrix)  and equating to zero, we have 
\begin{equation*}
  {\bf h}^\ast {\bf h}^T - {\bf D} + {\bf M} = 0.
\end{equation*}
Using the complementary slackness condition ${\bf M} {\bf Q} =0$, we
obtain
\begin{equation*}
{\bf D} {\bf Q} ={\bf h}^\ast {\bf h}^T {\bf Q}. 
\end{equation*}
Now ${\bf D}$ is full-rank because the shadow prices for increasing
antenna power are strictly positive. In other words, at optimum, the
power constraint must be met with equality, for otherwise we can
always increase the power and get a higher rate; hence the associated 
dual variables are strictly positive at optimum. Thus at optimum, we
have $\text{rank}({\bf Q}^\star) \leq \text{rank}({\bf h})$.

\subsection{Optimal transmit covariance for Rayleigh fading channels}
\label{app_fading}
The optimal ${\bf Q}^\star$ for (\ref{erg_cap_peakpwr}) also must have the 
diagonal values $q_{ii} = P_i$, for otherwise, we can singularly
increase the diagonal value that is lower than $P_i$ to be equal $P_i$ 
and hence increase the instantaneous as well as the ergodic
capacity. The remaining question is to find the off-diagonal values
$q_{ij}$ ($i \neq j)$.

To solve problem (\ref{erg_cap_peakpwr}), we will first illustrate the
technique by solving the special case $n=2$, then generalize to any
$n$. For $n=2$, we need to find the off-diagonal value of ${\bf Q}$, which
are $q_{21} = q_{12}^\ast$. Denote $q = q_{21}$, the problem becomes
\begin{align*}
  \max & \;\;\; E_{h_1,h_2} \left[ \log\left(P_1 |h_1|^2 + P_2|h_2|^2 +
      q^\ast h_1h_2^\ast + q h_1^\ast h_2 \right) \right]\\
  \text{s.t.} & \;\;\; |q|^2 \leq P_1P_2. \nonumber
\end{align*}
Let $J$ denote the objective function. Noting that $h_1$ and $h_2$ are
i.i.d. and complex Gaussian with zero-mean, then $-h_1$ also has the
same complex Gaussian distributions and is independent of $h_2$. Thus
flipping the sign of $h_1$ does not change the objective function, and
we can write
\begin{align*}
  J & = E_{h_1,h_2} \left[ \log\left(P_1 |h_1|^2 + P_2|h_2|^2 +
      q^\ast h_1h_2^\ast + q h_1^\ast h_2 \right) \right]\\
    & = E_{h_1,h_2} \left[ \log\left(P_1 |h_1|^2 + P_2|h_2|^2 -
      q^\ast h_1h_2^\ast - q h_1^\ast h_2 \right) \right]\\
    & = \frac{1}{2} E_{h_1,h_2} \left[\log \left\{ \left(P_1 |h_1|^2 +
          P_2|h_2|^2 + q^\ast h_1h_2^\ast + q h_1^\ast h_2 \right)
      \right. \right. \\ 
    & \quad\quad\quad\quad\quad\quad
    \left. \left. \times \left(P_1 |h_1|^2 + P_2|h_2|^2 - q^\ast
        h_1h_2^\ast - q h_1^\ast h_2 \right) \right\} \right]\\
    & = \frac{1}{2} E_{h_1,h_2} \left[ \log\left\{ (P_1 |h_1|^2 +
        P_2|h_2|^2)^2 - (q^\ast h_1h_2^\ast + q h_1^\ast h_2)^2
      \right\} \right] \\
    & \leq E_{h_1,h_2} \left[ \log\left( P_1 |h_1|^2 + P_2|h_2|^2
      \right) \right],
\end{align*}
where equality occurs if and only if $q^\ast h_1h_2^\ast + q h_1^\ast
h_2 = 0$ for all $h_1, h_2$, which implies $q = 0$.

Thus because of the symmetry in the distribution of the Rayleigh
fading channel, the optimal input covariance ${\bf Q}$ is a diagonal matrix,
${\bf Q} = \text{diag}\{P_1,P_2\}$. The constraint on $q$ is not active.

In the general case of any $n$, the objective function in problem
(\ref{erg_cap_peakpwr}) can be expressed as
\begin{align*}
  J &= E_{{\bf h}} \left[ \log\left(1 + \frac{1}{\sigma^2} \sum_{i=1}^n
      P_i|h_i|^2 + \frac{1}{\sigma^2} \sum_{i \neq j} q_{ij}h_i^\ast
      h_j \right) \right].
\end{align*}
Again since $\{h_i\}$ are i.i.d. Gaussian with zero mean, $-h_i$ for
any particular $i$ is also i.i.d. with the rest of the channel
coefficients. Thus we can successively flip the sign of a different
channel coefficient $h_i$ at each time, each time resulting in $q_{ij}
= q_{ji}^\ast = 0$ for all $j \neq i$ for $J$ to be maximized. Hence
the maximum value of $J$ is achieved when $q_{ij}=0$ for any $i \neq
j$. Therefore the optimal input covariance is diagonal, ${\bf Q} =
\text{diag}\{P_1, \ldots, P_n\}$.

\fontsize{11}{11}
\selectfont
\bibliographystyle{IEEEtran}
\bibliography{../../../reflist}

\end{document}